\documentclass[a4paper,11pt]{article}
\usepackage{jinstpub} % for details on the use of the package, please see the JINST-author-manual
\usepackage{lineno}

\usepackage{comment,mfirstuc}
\usepackage{enumitem}
\usepackage{hyperref}
\usepackage{graphicx}
\usepackage{caption}
\usepackage{subcaption}
\usepackage{multirow}
\usepackage{tabularx}
\usepackage{xspace}

\usepackage{hyperref}

\usepackage{tikz}
\usetikzlibrary{shapes,arrows,positioning}

\usepackage{comment}
\usepackage{caption}
\usepackage{subcaption}

\title{\boldmath Scalable AI-assisted Workflow Management for Detector Design Optimization Using Distributed Computing}

\author[4]{D. Anderson}
\author[1]{A. Bashyal}
\author[4]{M. Diefenthaler}
\author[5]{C. Fanelli}
\author[1]{W. Guan}
\author[2]{T. Horn}
\author[1]{A. Jentsch}
\author[1]{M. Lin}
\author[1]{T. Maeno}
\author[3]{K. Nagai}
\author[5]{H. Nayak}
\author[3]{C. Pecar}
\author[5]{K. Suresh}
\author[6]{F.Y. Tsai}
\author[3,4]{A. Vossen}
\author[1]{T. Wang}
\author[1]{T. Wenaus}
\author[]{(AID(2)E collaboration)}
\affiliation[1]{Brookhaven National Laboratory, Upton, New York, USA}
\affiliation[2]{The Catholic University of America, Washington, DC, USA}
\affiliation[3]{Duke University, Durham, North Carolina, USA}
\affiliation[4]{Thomas Jefferson National Accelerator Facility, Newport News, Virginia, USA}
\affiliation[5]{William \& Mary, Williamsburg, Virginia, USA}
\affiliation[6]{Stony Brook University, Stony Brook, New York, USA}

\emailAdd{wguan2@bnl.gov}

\abstract{
The Production and Distributed Analysis (PanDA) system, originally developed for the ATLAS experiment at the CERN Large Hadron Collider, has evolved into a robust platform for orchestrating large-scale workflows across distributed computing resources. Coupled with its intelligent Distributed Dispatch and Scheduling (iDDS) component, PanDA now supports AI/ML-driven workflows through a scalable, flexible workflow engine.

We present an AI-assisted framework for detector design optimization that integrates multi-objective Bayesian optimization with the PanDA–iDDS workflow engine to coordinate iterative simulations across heterogeneous resources. The framework addresses the challenge of exploring high-dimensional parameter spaces inherent in modern detector design.

We demonstrate the framework using benchmark problems and realistic studies of the ePIC / dRICH detector for the Electron–Ion Collider. Results show improved automation, scalability, and efficiency in multi-objective optimization. This work establishes a flexible and extensible paradigm for AI-driven detector design and other computationally intensive scientific applications.

}


\begin{document}
\maketitle
\flushbottom

\section{Introduction}

The Production and Distributed Analysis (PanDA) system~\cite{PanDA-CSBS,panda1} is a mature workload management platform designed to handle large-scale data processing across heterogeneous and geographically distributed computing resources. By abstracting the underlying infrastructure, PanDA provides users with a unified interface for workload submission and management, enabling transparent and efficient utilization of distributed environments without requiring detailed knowledge of resource configurations.

The intelligent Distributed Dispatch and Scheduling (iDDS)~\cite{idds-epjc, idds} extends PanDA with advanced workflow orchestration capabilities for complex and dynamic applications. iDDS supports Directed Acyclic Graph (DAG)-based workflows, conditional execution, iterative processing, and polymorphic workloads, allowing flexible expression and automation of sophisticated computational pipelines.

\begin{figure}[h]
\centering
\includegraphics[width=12cm]{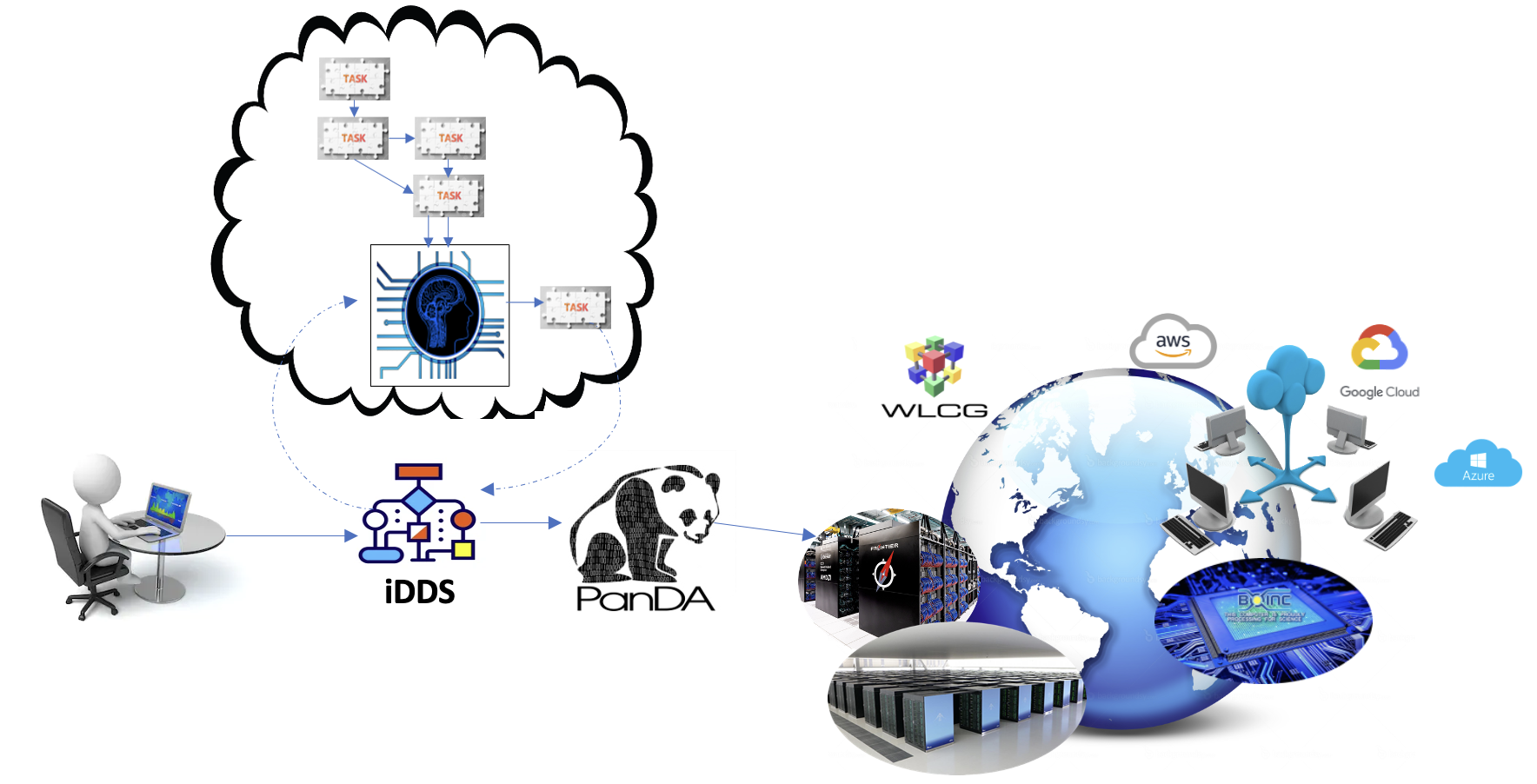}
\caption{An integrated workflow with PanDA and iDDS, where iDDS automates complex and dynamic workflows, and PanDA schedules workloads to large-scale distributed heterogeneous computing resources.}
\label{fig_ml_panda_idds}       % Give a unique label
\end{figure}

Together, PanDA and iDDS form a robust framework for distributed workflow management, widely adopted in production systems for large-scale data processing and analysis, as shown in Fig.~\ref{fig_ml_panda_idds}. Notable applications include the ATLAS~\cite{atlas} experiment at the Large Hadron Collider~\cite{lhc} and the Vera C. Rubin Observatory~\cite{rubin,rubin1}, where they support diverse workloads such as data carousels, hyperparameter optimization, active learning, and DAG-based data processing pipelines~\cite{idds_distributed_learning}.

Detector design represents a particularly demanding application domain, characterized by high-dimensional parameter spaces, complex constraints, and competing optimization objectives. Performance metrics such as spatial and momentum resolution, particle identification efficiency, and geometric acceptance must be evaluated through computationally intensive simulations, making exhaustive parameter exploration infeasible. Recent advances in machine learning, particularly multi-objective Bayesian optimization, offer powerful strategies for exploring these high-dimensional design spaces efficiently. However, integrating ML-driven optimization with distributed computing infrastructures remains a significant challenge, especially when coordinating large numbers of simulation and reconstruction tasks.

We present a scalable framework developed within the AI-assisted Detector Design for the Electron–Ion Collider (AID(2)E) project~\cite{aid2e, Diefenthaler_2024}. The AID(2)E framework is structured around three key components: flexible detector configurations, AI-driven optimization, and adaptable execution backends. The detector configuration layer is designed to be extensible, supporting a wide range of use cases—from specific detector systems such as the dual-radiator Ring Imaging Cherenkov (dRICH) and Barrel Imaging Calorimeter (BIC) to more general detector design problems. On top of this, AID(2)E explores multiple AI-driven optimization strategies, including multi-objective Bayesian optimization (MOBO) and multi-objective genetic optimization (MOGO), to efficiently navigate high-dimensional design spaces and balance competing objectives.

To support diverse computational environments, the framework integrates multiple execution backends. A local \texttt{joblib}-based runner enables rapid prototyping on a single machine, while a SLURM-based runner supports scalable execution on clusters and high-performance computing systems. For large-scale, geographically distributed workloads, a PanDA-based runner integrates with the iDDS workflow orchestrator to coordinate execution across heterogeneous resources. This multi-runner design enables seamless portability of workflows across local, HPC, and distributed infrastructures.

In this work, we focus on the PanDA-based execution model, which represents the most complex and scalable deployment scenario. We investigate the behavior of the PanDA scheduler and iDDS workflow orchestration under distributed conditions, including task concurrency, scheduling overhead, and system scalability. By coupling AI-driven optimization with distributed workflow management, the framework enables automated, end-to-end optimization across heterogeneous resources, providing a flexible and extensible paradigm for large-scale detector design and other computationally intensive scientific applications.

\section{Distributed workflow orchestration}

In this section, we present an overview of the distributed workflow architecture underlying the AID(2)E framework. We begin by describing the system design that integrates AI-driven optimization with the PanDA/iDDS infrastructure. We then outline the role of PanDA in large-scale workload management and the Function-as-a-Task paradigm in iDDS for workflow orchestration. Finally, we describe how these components are combined to enable scalable, AI-assisted detector design.

\paragraph{Architecture}

\begin{figure}[!ht]
    \centering
    \includegraphics[scale = 0.5]{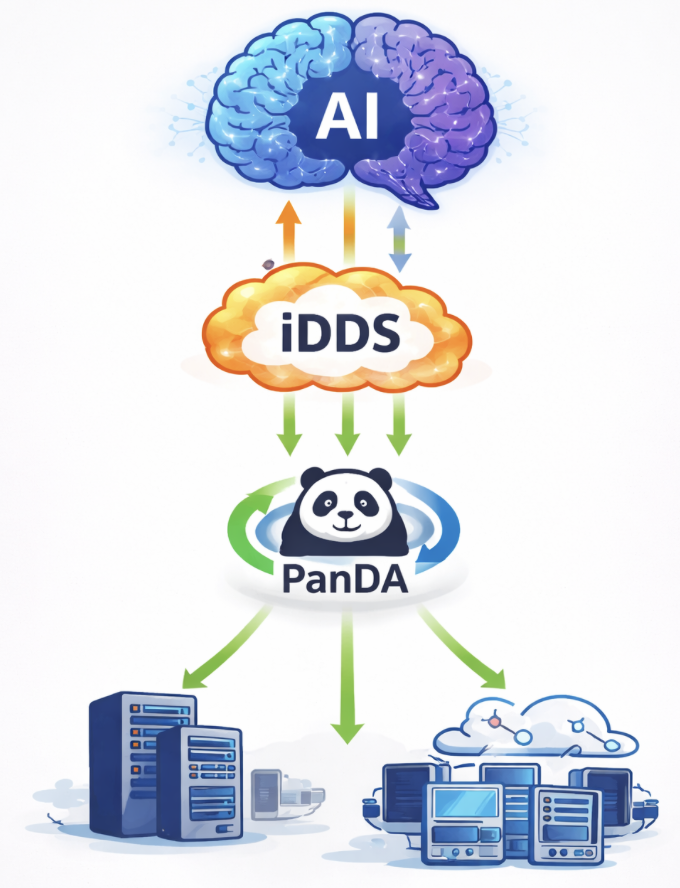}
    \caption{\textbf{AI integration with PanDA/iDDS.} iDDS maps AI pipeline functions to remote tasks executed by PanDA across distributed resources and asynchronously aggregates results, enabling a workflow that behaves like local function execution.}
    \label{fig:ai_idds_panda}
\end{figure}

As the scope and complexity of scientific workflows continue to grow, managing intricate dependencies and execution logic becomes increasingly challenging. In particular, adapting AI pipelines to offload selected components as remotely executed tasks often requires significant restructuring, making the process cumbersome and error-prone. 

To address this challenge, iDDS introduces a Python-based abstraction that leverages decorators to transform local functions into distributed PanDA workloads. This approach enables users to define workflows and dependencies using familiar programming constructs, while seamlessly executing selected components across distributed resources with minimal modification to the original AI pipeline. In addition, iDDS asynchronously aggregates the resulting outputs, allowing computationally intensive workloads to be transparently offloaded to remote resources. This design provides a high-level abstraction that encapsulates distributed execution, enabling the workflow to exhibit semantics analogous to local function execution from the user’s perspective.

% As illustrated in Fig.~\ref{fig:ai_idds_panda}, the overall architecture integrates three key components. The AI optimization layer, based on Bayesian optimization with Ax/BoTorch, generates candidate detector configurations and updates surrogate models as new results become available, guiding efficient exploration of the design space. The iDDS system serves as the workflow orchestrator, transforming user-defined Python functions into distributed tasks, handling code packaging and deployment, enabling parallel execution across parameter sets, and supporting asynchronous result retrieval. Finally, PanDA provides the underlying workload management layer, executing tasks across heterogeneous resources, including grid, cloud, HPC, and Kubernetes platforms, thereby enabling scalable optimization campaigns.

The architecture combines three layers (Fig. ~\ref{fig:ai_idds_panda}): (1) an AI optimization layer using Ax/BoTorch62 for candidate generation and surrogate model updates; (2) iDDS as workflow orchestrator, handling packaging, deployment, and parallel execution; and (3) PanDA as the workload manager across grid, cloud, HPC, and Kubernetes platforms.

\paragraph{Distributed workload management}

PanDA provides a unified workload management framework for large-scale distributed computing. It offers a consistent interface through which users from different institutions can submit and manage jobs via HTTP-based services, supported by common authentication mechanisms such as X.509 certificates or OpenID Connect~\cite{oidc}. PanDA integrates heterogeneous computing resources—including grid, cloud, Kubernetes, and high-performance computing systems—while abstracting site-specific differences in software stacks and schedulers such as Slurm, HTCondor, and PBS. This abstraction significantly reduces user complexity, enabling seamless execution across geographically distributed resources. Proven at scale in the ATLAS experiment at the Large Hadron Collider, where it serves thousands of users across more than 170 sites, and in the Rubin Observatory, PanDA has demonstrated robust performance and scalability, and is now being extended to support emerging use cases at the Electron–Ion Collider.

\paragraph{Workflow orchestration}

%The iDDS system introduces a Python-based abstraction that leverages decorators to transform local functions into distributed PanDA workloads through a Function-as-a-Task paradigm. In this model, user-defined Python functions are packaged and executed as distributed tasks, while results are returned asynchronously to the driving workflow. This approach enables complex workflows to be expressed using standard Python, while seamlessly scaling execution across heterogeneous computing resources.

% The Function-as-a-Task scheme provides a simplified and flexible interface for workflow definition. Users write standard Python programs and annotate selected functions with decorators to indicate remote execution, allowing intuitive expression of workflow logic without requiring explicit management of distributed infrastructure.

iDDS employs the Function-as-a-Task paradigm to orchestrate workflows, comprising three main steps. First, user source code and execution context are packaged and uploaded to a cache service, from which a function wrapper initializes the runtime environment on remote resources; this environment may be container-based, Conda-based~\cite{conda}, or a standard Python installation. Second, decorated functions and their parameters are serialized into PanDA jobs, which are scheduled and executed on distributed resources, where the wrapper reconstructs and runs the original functions. Third, outputs are asynchronously returned to the caller, enabling non-blocking execution and iterative processing.

Result retrieval supports both STOMP~\cite{stomp}- and HTTP REST~\cite{rest}-based communication, with STOMP via ActiveMQ~\cite{activemq} as the default for efficient transfer and REST as a fallback for robustness across environments.

This paradigm allows transparent execution of user-defined functions on distributed resources through PanDA. Leveraging existing infrastructure, it requires minimal configuration while achieving high scalability. Asynchronous, publish–subscribe-based result handling further decouples task execution from workflow control, making the approach well suited for large-scale, AI-driven optimization workflows.

\paragraph{AID(2)E workflow}

\begin{figure}[h]
    \centering
    \begin{subfigure}[b]{0.49\textwidth}
        \centering
        \includegraphics[width=\textwidth]{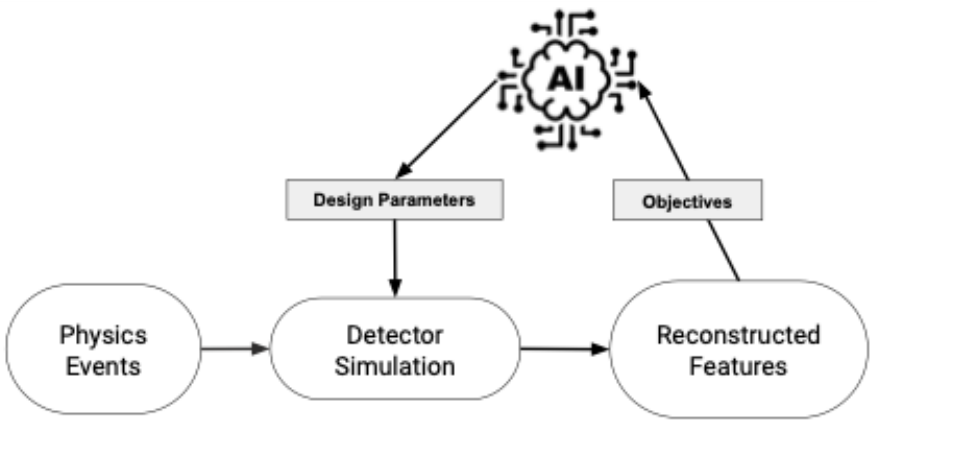}
        \caption{Baseline AID(2)E workflow}
        \label{fig_aid2e}
    \end{subfigure}
    \hfill
    \begin{subfigure}[b]{0.49\textwidth}
        \centering
        \includegraphics[width=\textwidth]{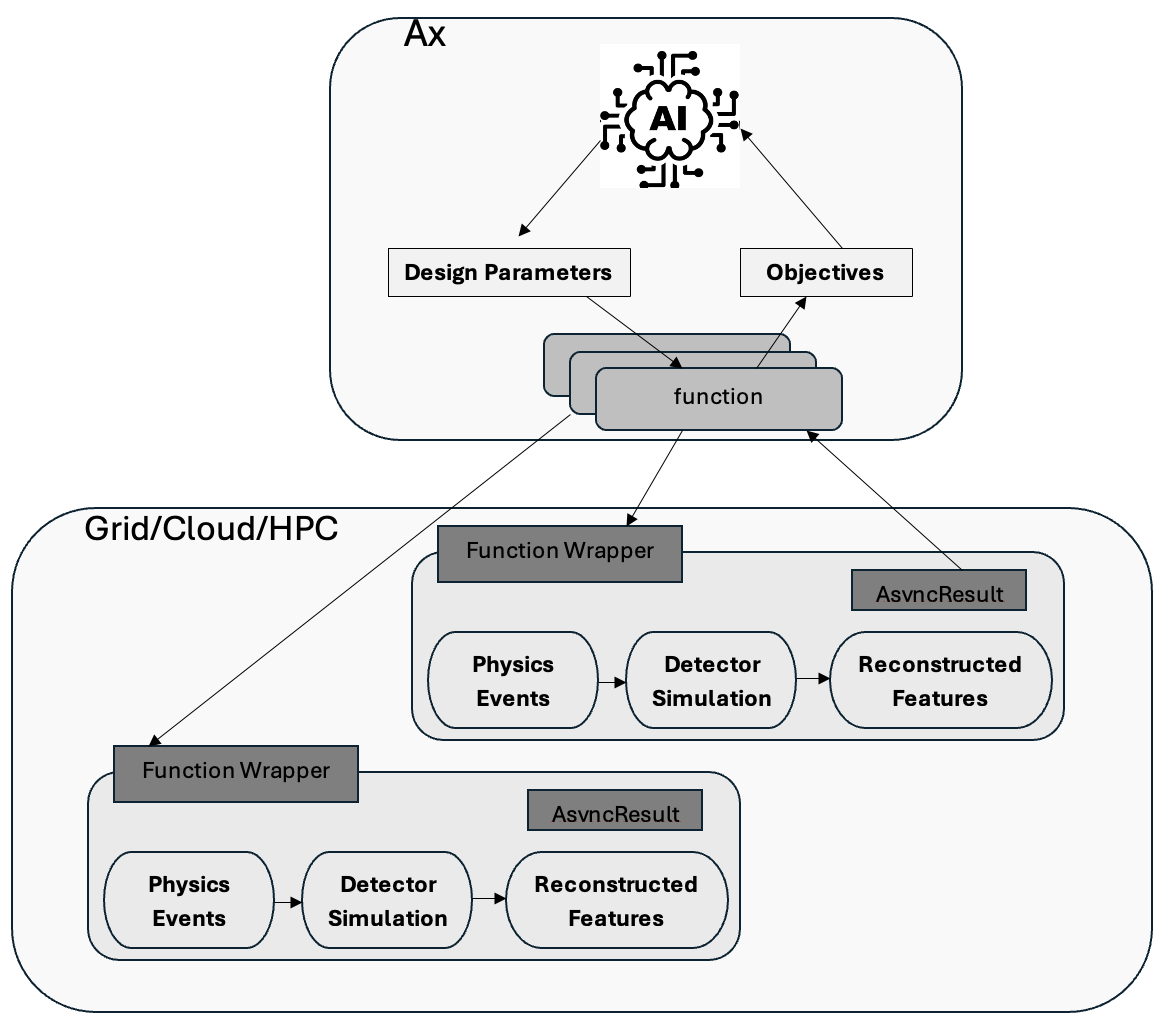}
        \caption{AID(2)E with Function-as-a-Task}
        \label{fig_aid2e_function_as_a_task}
    \end{subfigure}
    \caption{\textbf{AID(2)E workflow.} (a) The AI-driven framework proposes detector design parameters for multiple objectives, which are evaluated through simulation. (b) With the Function-as-a-Task paradigm, local functions are transformed into PanDA jobs and executed on distributed resources, enabling scalable and transparent workflow execution.}
    \label{fig:function}
\end{figure}

%The AID(2)E framework addresses limitations of traditional detector design optimization, which typically relies on subsystem-level tuning under global constraints and ad hoc or brute-force methods such as grid search. These approaches scale poorly with increasing parameter dimensionality, lack support for multi-objective optimization, and fail to capture global correlations across the full detector system. In contrast, detector design inherently involves competing objectives—such as performance, resolution, and cost—requiring exploration of a spectrum of trade-off solutions rather than a single optimum. Multi-objective optimization provides a principled approach to this problem, enabling efficient exploration of high-dimensional design spaces and revealing non-trivial correlations among parameters. Previous studies~\cite{aid2e} have shown that AI-assisted methods can outperform baseline designs across large parameter spaces.

%A key requirement of this workflow is the efficient coordination of optimization, simulation, and analysis across distributed computing resources. Evaluating large numbers of detector configurations demands high throughput and scalable workflow management. To this end, AID(2)E leverages the PanDA workload management system, orchestrated by iDDS, both of which have demonstrated scalability in large scientific applications.

The AID(2)E framework addresses the limitations of traditional detector design methods, which often rely on subsystem-level tuning or brute-force approaches like grid search. It implements multi-objective optimization to efficiently explore competing goals—such as performance, resolution, and cost—while uncovering complex correlations in high-dimensional parameter spaces. By integrating machine learning with realistic simulation pipelines, AID(2)E enables AI-driven, system-level optimization of the full detector.

The workflow leverages PanDA and iDDS to distribute computationally intensive simulation and reconstruction tasks across heterogeneous resources. Results are returned asynchronously to guide iterative, adaptive optimization, allowing scalable and automated exploration of complex detector designs. This combination of AI and distributed computing ensures efficient resource utilization while controlling computational cost.

\section{Experiments}

\begin{figure}[!ht]
    \centering
    \includegraphics[trim={0cm 5cm 0cm 0cm},clip,width=0.95\textwidth]{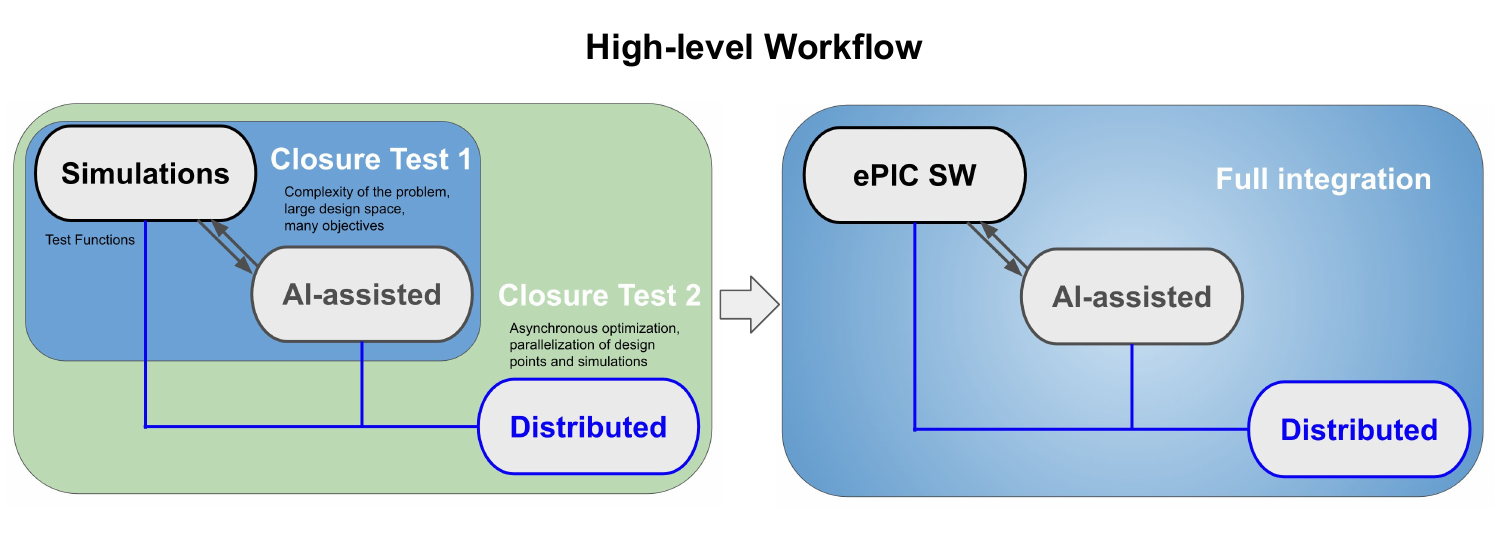}
    \caption{\textbf{High-level workflow of AID(2)E.} The staged validation strategy consists of three steps: closure test~1 validates the optimization algorithms using benchmark problems with known Pareto fronts; closure test~2 evaluates distributed workflow execution across heterogeneous resources using PanDA/iDDS; and the full integration stage applies the framework to compute-intensive ePIC simulation and reconstruction tasks.}
    \label{fig:proposal_workflow}
\end{figure}

The AID(2)E project follows a staged closure-test strategy to validate the complete AI-assisted detector design workflow. Closure test~1 focuses on optimizer validation, using benchmark problems with known Pareto fronts to verify the convergence and performance of multi-objective optimization algorithms. Closure test~2 focuses on distributed workflow execution, evaluating the scalability and efficiency of PanDA/iDDS-based orchestration across heterogeneous computing resources. The final integration stage replaces the benchmark objectives with compute-intensive ePIC detector simulation and reconstruction workflows.

This work focuses on closure test~2, where the optimization methodology is coupled with distributed workflow infrastructure. Results from closure test~1 are not presented here because the goal of this paper is to evaluate the distributed execution capabilities required for realistic detector optimization workloads, while the optimizer validation has been established independently using standard benchmark problems.

% The AID(2)E project employs a sequence of closure tests to validate both the optimization methodology and the distributed workflow infrastructure. Closure test~1 verifies the convergence of multi-objective optimization algorithms using benchmark problems with known Pareto fronts, establishing a baseline for performance. This work focuses on closure test~2, which evaluates the scalability and efficiency of distributed workflow execution.

%Its primary objective is to demonstrate that large numbers of detector design evaluations can be efficiently executed across heterogeneous computing resources. In this stage, the optimization workflow is integrated with the PanDA workload management system, orchestrated by iDDS, enabling distributed execution of simulation tasks and asynchronous aggregation of results. A high-level overview of the framework is shown in Fig.~\ref{fig:proposal_workflow}.

\paragraph{DTLZ2 benchmark}

\begin{figure}[ht]
     \centering
     \begin{subfigure}[b]{0.48\textwidth}
         \centering
         \includegraphics[width=\textwidth]{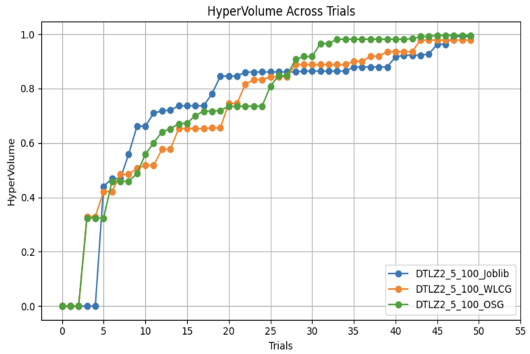}
         \caption{Hypervolume vs. trials}
         \label{fig_dtlz2}
     \end{subfigure}
     \hfill
     \begin{subfigure}[b]{0.48\textwidth}
         \centering
         \includegraphics[width=\textwidth, height=0.67\textwidth]{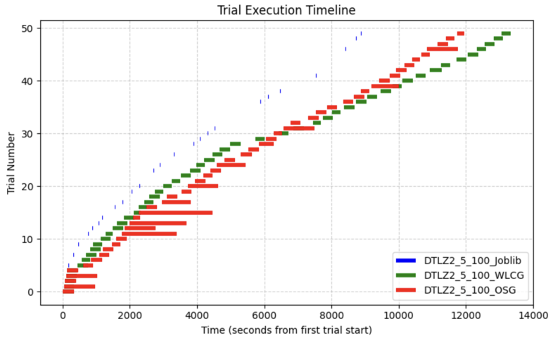}
         \caption{Hypervolume vs. trial execution time}
         \label{fig_dtlz2_time}
     \end{subfigure}
     \caption{DTLZ2 benchmark results for 5 objectives and 100 parameters, with \texttt{max\_concurrent\_trials}=5. The benchmark evaluates multiple execution backends: a local backend based on jobLib for single-resource execution, the Worldwide LHC Computing Grid (WLCG) backend, and the Open Science Grid (OSG) backend for distributed execution across heterogeneous resources. Distributed execution achieves similar hypervolume convergence while enabling asynchronous execution with up to five concurrent trials. The benchmark objective is computationally inexpensive; therefore, the measured runtime includes optimization overhead and workflow scheduling effects rather than being dominated solely by objective evaluation.}
     \label{fig_dtlz2_results}
\end{figure}

%\caption{DTLZ2 benchmark results for 5 objectives and 100 parameters, with \texttt{max_concurrent_trials}=5. Distributed execution achieves similar hypervolume convergence while enabling asynchronous execution with up to five concurrent trials. The benchmark objective is computationally inexpensive; therefore, the measured runtime includes optimization overhead and workflow scheduling effects rather than being dominated solely by objective evaluation.}

The DTLZ2~\cite{deb2006multi} problem is used in closure test~2 to evaluate the performance of the PanDA/iDDS system on distributed resources. Benchmarks were conducted across configurations with varying numbers of objectives and design parameters; here, we present a representative case with five objectives and 100 parameters.

This benchmark provides a controlled environment for validating distributed workflow orchestration prior to full detector simulations. As shown in Fig.~\ref{fig_dtlz2_results}, the two panels demonstrate that distributed execution preserves optimization quality while reducing the wall-clock time required to complete the optimization campaign through concurrent trial execution. In this benchmark, the maximum number of concurrent trials was configured to five, representing a controlled demonstration of asynchronous distributed execution rather than a limitation of the framework. Comparable hypervolume convergence is observed between local and distributed execution, while the runtime comparison highlights the benefit of parallel trial scheduling.

For this benchmark, the objective evaluation itself is inexpensive compared with realistic detector simulations, and therefore the measured runtime also includes optimization overhead and workflow scheduling effects. In realistic detector design workflows, where simulation and reconstruction dominate the trial cost, the same distributed execution strategy can provide a substantially larger benefit; with five concurrent simulations, the ideal throughput improvement can approach a factor of five when simulations are the dominant contribution to runtime. These results demonstrate that the PanDA/iDDS framework enables scalable optimization workflows across distributed resources without degrading optimization performance.

\paragraph{dRICH detector optimization}

To demonstrate applicability to realistic detector design, the workflow is applied to the dual-radiator Ring Imaging Cherenkov (dRICH) detector in the ePIC experiment as outlined in \cite{Diefenthaler_2024}. In this setup, multi-objective Bayesian optimization is coupled with PanDA/iDDS to orchestrate detector simulation, objective evaluation, and result aggregation across distributed resources.

The optimization targets key physics performance metrics, including pion–kaon and kaon–proton separation and detector acceptance over a broad momentum range. The design space consists of seven parameters, including aerogel geometry, mirror configuration, and photosensor placement, subject to geometric constraints and overlap checks.

\begin{figure}[!ht]
    \centering
    \includegraphics[height=0.55\textwidth,width=0.75\textwidth]{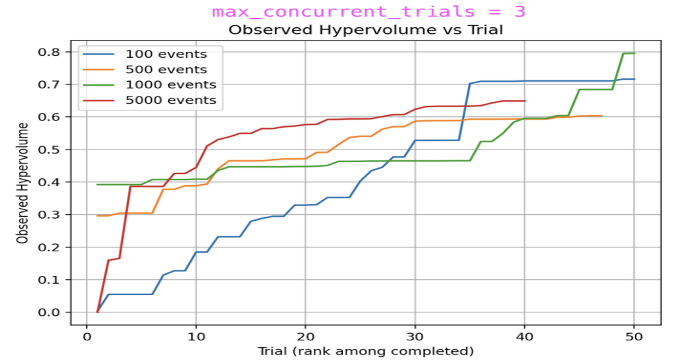}
    \caption{\textbf{dRICH multi-objective optimization.} Hypervolume as a function of optimization trials for the distributed workflow, demonstrating progressive improvement of detector configurations.}
    \label{fig:drich_mobo}
\end{figure}

As shown in Fig.~\ref{fig:drich_mobo}, the hypervolume increases monotonically with the number of trials, indicating continuous improvement of detector configurations. The different event-count cases (100, 500, 1000, and 5000 simulated events per trial) illustrate the workflow behavior as the computational cost of individual simulation evaluations increases, demonstrating the ability of the distributed framework to maintain optimization progress under varying resource demands. The optimization progressively identifies detector configurations with improved trade-offs among the target physics objectives, leading to enhanced combined particle-identification performance through improved pion--kaon and kaon--proton separation while maintaining detector acceptance requirements. The distributed workflow enables efficient scaling across resources, improving throughput and automation while maintaining robustness through asynchronous execution.

These results demonstrate that the AID(2)E framework effectively integrates AI-driven optimization with distributed computing infrastructure, enabling scalable and automated detector design for realistic experimental systems.

\section{Discussion and Future Work}

Performance evaluation demonstrates that the PanDA/iDDS framework scales effectively with concurrent execution, while also highlighting overheads from trial generation, scheduling latency, and resource contention in distributed environments. Asynchronous execution and parallel evaluation substantially improve efficiency, particularly for computationally intensive simulations, making the framework most advantageous when evaluation costs dominate optimizer overhead—typical of realistic detector-design studies.

The integration of AI-driven optimization with distributed computing enables efficient exploration of high-dimensional parameter spaces, reduces time-to-solution, and enhances automation, reproducibility, and overall workflow robustness. By coupling multi-objective optimization with scalable workflow orchestration, the framework provides a practical solution for end-to-end optimization of complex detector systems.

Future work will focus on extending the framework to support larger machine learning models, additional detector components, and leveraging large language models for advanced workflow automation and decision support. Furthermore, ongoing developments aim to generalize the approach for broader applications and to extend the workflow scheduling infrastructure to SLURM-based runners, enabling AI-assisted optimization across other detectors and experiments at the Electron–Ion Collider. This scalable, distributed, and AI-driven methodology promises to accelerate detector design cycles, uncover optimal trade-offs among competing objectives, and establish a foundation for future large-scale nuclear and particle physics experiments.

\appendix

\acknowledgments

C.F., K.S. and H.N. were supported by the Office of Nuclear Physics of the U.S. Department of Energy under Grant Contract No. DE-SC0024625. C.P. and K.N. were supported by the Office of Nuclear Physics of the U.S. Department of Energy under Grant Contract No. DE-SC0024478. M.D. was supported by the U.S. Department of Energy Office of Science, Office of Nuclear Physics contract number DE-AC05-06OR23177, under which Jefferson Science Associates, LLC operates Jefferson Lab. T.H. was supported by DOE U.S. Department of Energy Office of Science, Office of Nuclear Physics contract number DE-SC-0024691. This work was supported by the U.S. Department of Energy, Office of Science, under Brookhaven Science Associates (BSA) contract number DE-SC0012704.

%\paragraph{Note added.} This is also a good position for notes added after the paper has been written.

% Bibliography

\bibliography{biblio}

\end{document}